\documentclass{aa}  

\usepackage{graphicx}
\usepackage{txfonts}
\newcommand*\dd{\textrm{d}}

\begin{document} 

   \title{Optically thick, nonlocal, inhomogeneous, stationary jet model for high-energy radiation from blazars: Application to  Mrk 421}
   \titlerunning{Inhomogeneous model for blazars}

   \author{Piotr Banasinski
          \inst{}
          \and
          Wlodek Bednarek
          }

   \institute{Department of Astrophysics, Faculty of Physics and Applied Informatics,  University of Lodz,
              ul. Pomorska 149/153, 90-236 Lodz, Poland\\
              \email{pgbanasinski@gmail.com; wlodzimierz.bednarek@uni.lodz.pl}
             }

   \date{Received May 16, 2022; accepted Sep 22, 2022}
 
  \abstract
   {There is an increasing number of observational evidence that very high energy $\gamma$-rays in radio-loud activ galactic nuclei are produced in the direct vicinity of a supermassive black hole (SMBH), close to the base of a relativistic jet. In the case of some blazars, the angle between the jet axis and the observer's line of sight is smaller than the angular extent of the jet.  $\gamma$-rays that are produced close to SMBH therefore have to propagate in the nonthermal radiation 
of the extended jet before reaching the observer. This $\gamma$-ray emission can be strongly absorbed in the extended
jet radiation, producing a second generation of $e^\pm$ pairs that loses energy mainly via the synchrotron process.}
   {We developed a nonlocal, inhomogeneous, stationary jet model in order to describe the
multiwavelength emission from blazars. With this advanced model, we investigated the impact of the extended 
jet radiation on the propagation of $\gamma$-rays that are ejected from the direct vicinity of SMBH toward an observer located within the solid angle of the jet. We determined the conditions under which $\gamma$-rays are absorbed in the jet radiation and explored the effect of this absorption process on the $\gamma$-ray spectra and on the hard X-ray emission
observed from some blazars.}
   {We first developed an inhomogeneous, stationary jet model in which the radiation that is produced nonlocally in the
jet was taken into account when we calculated the nonthermal emission in the broad energy range. This emission serves
as a target on which $\gamma$-rays, produced close to SMBH, can be absorbed. 
As a result, the cascade is initiated within the jet through inverse Compton and synchrotron processes.}
   {We show that this advanced inhomogeneous jet model can explain the multiwavelength spectrum of the BL Lac object Mrk 421 in a nonflaring state for reasonable parameters of the jet and the SMBH. Moreover, we argue that synchrotron emission from the secondary $e^\pm$ pairs, which appear as a result
of absorption of $\gamma$-rays that are produced close to the SMBH within the jet radiation, is consistent with the concave hard X-ray emission observed from Mrk 421.}
   {}


   \maketitle
%

\section{Introduction}

Blazars belong to a class of active galaxies that contain a supermassive black hole (SMBH) and a jet, whose emission can vary on timescales as short as a few minutes \citep{Bland19}.  Some $\gamma$-ray flares are observed 
on a timescale corresponding to (or even shorter than) the light-crossing time of their SMBH. 
The most extreme examples, with a time variability of some minutes, are PKS 2155-304 \citep{Aharo07} and Mrk 501 \citep{Alber07}. This extremely short timescale variability may indicate that at least a part of the very high energy (VHE) $\gamma$-rays is produced in the close vicinity of their SMBHs. This explanation is especially favorable for BL Lacs, where radiation from the broad-line region and the accretion disk is less intense than in the case of flat-spectrum radio quasars. Therefore, VHE $\gamma$-rays, which are proposed to be produced close to the SMBH, may be able to avoid strong absorption \citep{Donea03}. 

The hypothesis that the VHE $\gamma$-rays are produced in the close vicinity of the SMBH is also supported by the 
$\gamma$-ray variability that is observed in the case of radio galaxies that are viewed at a relatively large angle to the observer's line of sight. Variability of TeV $\gamma$-ray emission on a timescale as short as two days has been observed from M 87 \citep{Aharo06}. An exceptionally short flare, on a timescale of 5 min, 
has also been observed from IC 310 \citep{Aleks14}. This flare was characterized by a very hard spectrum (differential spectral index flatter than -2), which extended up to $\sim$10 TeV. In the case of both objects, M 87 and IC 310, it is unlikely that the emission was produced within the extended jet because the observation angle is relatively large.

A number of models have been proposed in order to explain the production of this extreme VHE $\gamma$-ray emission in the vicinity of an SMBH. The first group of models, inspired by the pulsar models, proposes the existence of a vacuum gap in the SMBH magnetosphere in which particles can reach extreme energies \citep[see, e.g.,][]{Neron07, Hirot16, Kisak22}. Recently, these models were used to explain the narrow feature in the VHE $\gamma$-ray spectrum of Mrk 501 \citep{MAGIC2020, Wende21}. The second group of models assumes that the electric field is generated in the reconnection process of the magnetic field in the accretion disk or the disk corona.  Electrons can be accelerated to VHE energies in this electric field \citep{Bedna97}. Other models assumed that the emission is produced in the interaction of the jet with compact objects, such as luminous or late-type stars 
\citep{Bedna97b, Barko12}. Another model, called the jets-in-jet model \citep{Giann09},  indicates that the magnetic reconnection process can occur close to the base of the Poynting-dominated jet. 

It is interesting to investigate the consequences of the appearance of VHE $\gamma$-rays in blazars whose
jets are inclined at angles smaller than the opening angle of the jet itself. Then, the VHE $\gamma$-rays that are produced in the vicinity of the SMBH have to pass through the radiation field of the extended jet before reaching the observer. These $\gamma$-rays can be absorbed in collisions with the soft, nonthermal radiation from the jet. A part
of the VHE emission is expected to initiate the synchrotron inverse Compton (IC) $e^\pm$ pair cascade inside the jet.
These processes are the main topic of our investigations.
We analyze the effects of absorption of the VHE $\gamma$-rays in the radiation field of the extended jet assuming that they are produced close to the jet base. Absorbed $\gamma$-rays are converted into secondary $e^\pm$ pairs, which lose energy in the IC and synchrotron processes.

This article investigates two tasks. In the first task, we develop the nonlocal, inhomogeneous, stationary jet model (Section 2). We use this model to calculate the multiwavelength spectrum of the BL Lac object Mrk 421 in the nonflaring state (Section 3). In the second task, we develop a numerical model of the electromagnetic cascades initiated in the radiation of the inhomogeneous jet (Section 4). The model is adopted to explain the concave shape of the hard X-ray emission observed from  Mrk 421 (Section 5). The conclusions are presented at the end of this paper in Section 6.

%
\section{Nonlocal inhomogeneous stationary jet}

At first approximation, the variable nonthermal emission from blazars is usually interpreted in terms of the homogeneous synchrotron self-Compton (SSC) model. This model assumes that a compact emission region (blob), with fixed parameters, moves along the jet with a large Lorentz factor. This simple homogeneous model can naturally explain the relatively short-timescale flares that are observed from blazars provided that the blob is small enough and/or its Lorentz factor is large. However, the long-timescale persistent emission from blazars cannot be described by this simple model. It is expected that during the motion of the blob along with the jet, 
the basic physical parameters responsible for radiation processes change significantly. For a jet pointed at the observer, the distance traveled by the blob can be estimated with $d = 2 \Gamma_j^2 \beta c \Delta t$, where $\beta c$ is the velocity of the blob, $\Delta t$ is the observed timescale, and $\Gamma_j$ is the Lorentz factor of the blob. For a typical observation time of stationary emission from the blazar of about a month ($\Delta t = 2.6 \times 10^6 $ s) and the Lorentz factor of the jet ($\Gamma_j \sim 10$), the distance traveled by the emission region is about a few parsecs. Therefore, the basic parameters of the emission region (e.g., the radius and the magnetic field strength) has to change significantly. 

Steady nonthermal emission, with a timescale greater than a month, was observed from two of the best investigated BL Lacs so far, that is, Mrk 421 \citep{Abdo11a, Aleks15a, Accia21} and Mrk 501 \citep{Abdo11b, Aleks15b}. During multiwavelength campaigns dedicated to these sources, both BL Lacs were observed in the low-activity state. For most of these observations, Mrk 421 and Mrk 501 emitted low-level, persistent nonthermal radiation in a low state, without clear flaring events \citep{Ander09, Accia21}. 

In order to describe this long-term and persistent nonthermal emission, we propose a scenario in which the majority of radiation is produced in the extended (parsec-scale) part of a jet. To describe this emission, we developed an inhomogeneous and stationary jet model, which, in contrast to previous studies, also takes into account the effects of 
nonlocally produced radiation in other parts of the jet on relativistic electrons that are located in a specific region of the jet. In this model, the basic parameters of the jet, such as the strength of the magnetic field and the spectrum of electrons, vary along the jet. Due to the elongated shape of the jet, the radiation produced by electrons in the specific parts of the jet is observed as strongly anisotropic by electrons located in a specific region of the jet.
This is also true for the observer in the reference frame comoving with the jet. In our version of the inhomogeneous jet model, the IC scattering of nonlocally produced photons therefore also has to be taken into account when
considering emission from relativistic electrons.

Different versions of the inhomogeneous models have been investigated up to now. \citet{Bland79} proposed a model of a conical inhomogeneous jet in which the magnetic field decreased linearly with the distance from the jet base and the electron density was in equipartition with the energy density of the magnetic field. They showed that the flux of the radio emission from a jet like this is flat. The other inhomogeneous models also attempted to explain the multiwavelength emission from active galactic nucelus (AGN) jets \citep[see, e.g.,][]{Ghise85, Boute08, Potte12}. It is important to note that the inhomogeneous models were designed for black hole binaries \citep[e.g.,][]{Zdzia14}. Some of the inhomogeneous models focused on explaining the time structure of the emission. For example, \citet{Asano14} considered thin shells propagating along with the jet, whereas \citet{Katar01} proposed a blob-in-jet model in which a compact blob moves inside the stationary jet. Recently, \citet{Boula22} developed a model of an expanding blob in order to explain the temporal behavior of the flare event.

\begin{figure}
  \resizebox{\hsize}{!}{\includegraphics{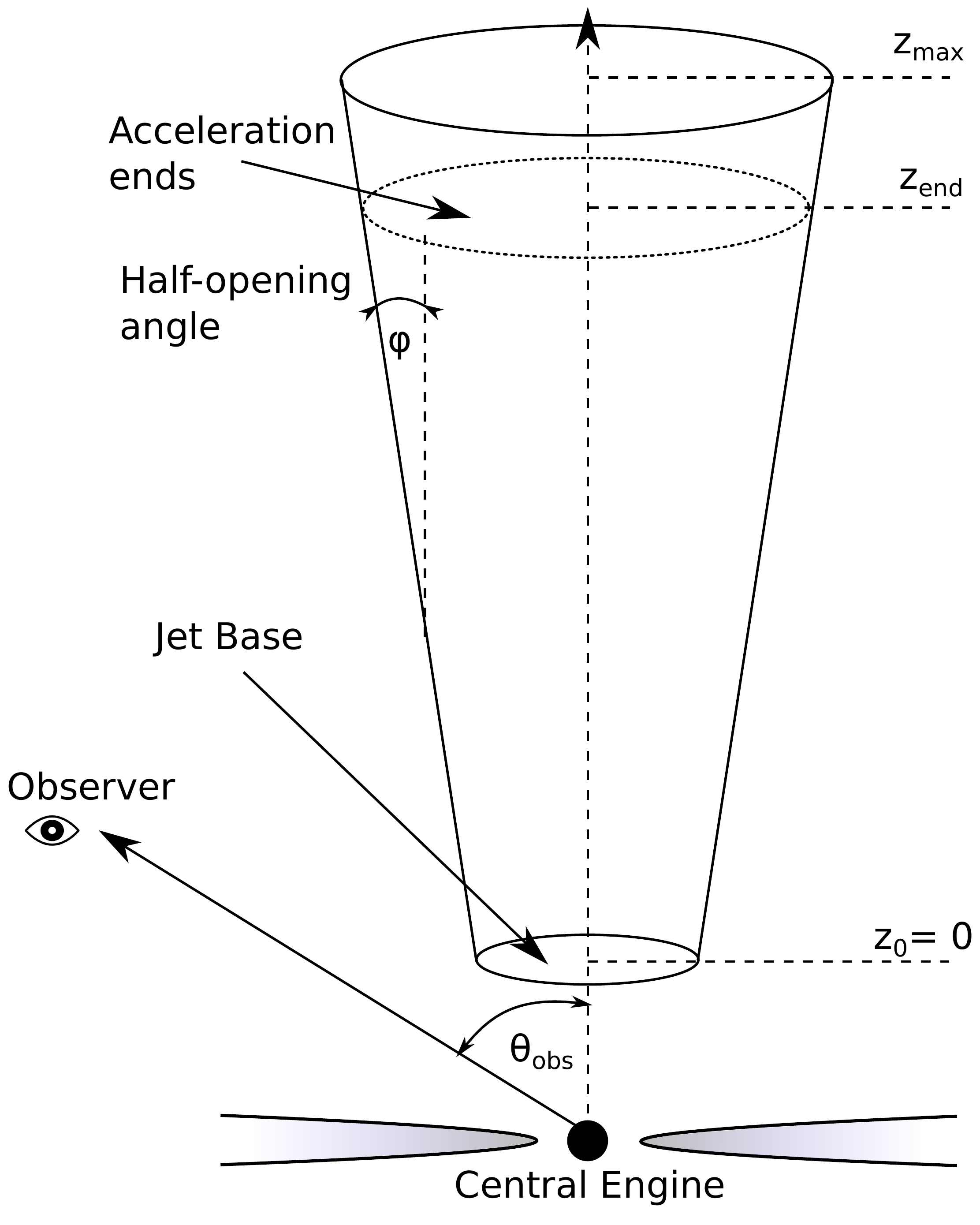}}
  \caption{Schematic picture (not to scale) of the nonlocal inhomogeneous jet as seen by an external observer. The jet has a truncated conical shape in a range of distances from the SMBH between $z_0 = 0$ and $z_{max}$. The inner radius of the jet at the jet base is $R_0$. Electrons are accelerated in the part of the jet between $z_0$ and $z_{end}$. We follow their cooling in different processes up to $z_{max} = 10 z_{end}$. The half-opening angle of the jet is equal to $\phi$. The jet is observed at an angle $\theta_{obs}$ that can be larger or smaller than $\phi$.}
  \label{jet-scheme}
\end{figure}

\subsection{Description of the jet model}

We approximated the geometry of the jet by a cone truncated from both sides (see Fig.~\ref{jet-scheme}). The base of the jet was located at a distance $z_{0}$. The radius at the jet base was fixed on $R_0 = 10 R_{Sch}$, where $R_{Sch}$ is the Schwarzschild radius of the SMBH. The radius of the jet increased with the distance $z$ along the jet according to $R_{obs}(z) = R_0 + z \tan \phi$, where $\phi$ is the half-opening angle of the jet. Because the typical value of the half-opening angle for relativistic jets is small, we used the approximate relation $R_{obs}(z) \simeq R_0 + z \phi$.
It is more convenient to consider the processes inside the jet in its plasma rest frame (the jet frame). In the jet frame, the geometry of the jet is described by $R_j(t') = R_0 + \phi \Gamma_j \beta_j c t'$, where $\beta_j c$ is the velocity of the plasma in the jet, and $t'$ is the time flow in the jet frame. The magnetic field in the jet was
assumed to drop inversely proportional to the radius of the jet, $B(t') = B_0 (R_0/R_j(t'))$, where $B_0$ is
the strength of the magnetic field at the jet base. This relation corresponds to the dominance of the toroidal component of the magnetic field \citep{Begel84}.

It is assumed that electrons are accelerated continuously in the active part of the jet between $z_0$ and $z_{\rm end}$. 
We do not concentrate on the details of the acceleration process. The acceleration process could be a consequence of either the shock waves (first-order Fermi mechanism, as postulated in the internal shock models) or of the magnetic field irregularities in the jet plasma (second-order Fermi mechanism). The spectrum of electrons was assumed to have a power-law type with an exponential cutoff at energy $E_{max} = \gamma_{max} mc^2$, where $\gamma_{max}$ is the characteristic maximum Lorentz factor of the electrons. Then, the spectrum of relativistic electrons in the jet, $Q_{inj}(\gamma, t') = \dd N / \dd z' \dd t' \dd \gamma$ , is described by
\begin{equation}
Q_{inj}(\gamma, t') = Q_0 \times \left( \frac{R_0}{R(t')} \right)^q \times \gamma^s \times \exp\left(-\gamma / \gamma_{max}(t')\right),
\end{equation}
\noindent 
where $s$ is the spectral index of the injected electrons, $Q_0$ is the normalization factor of the injected electrons at the jet base, and $q$ describes the injection profile along with the jet. In this modeling, the distribution of injected electrons is uniform for $q=0$. For $q=2$, the distribution of the relativistic electrons is proportional to the energy density of the toroidal magnetic field along with the jet. The total power in the relativistic electrons can be calculated from
\begin{equation}
L'_{inj} = m_e c^2 \int \int Q_{inj}(\gamma, t') \gamma \dd z' \dd \gamma.
\end{equation}
The maximum Lorentz factor of the injected electrons, $\gamma_{max}$, was obtained from the comparison of the acceleration timescale, $t'_{acc}$, with the total cooling timescale, $t'_{cool}$. The acceleration timescale, $t'_{acc}$, is described by the so-called acceleration coefficient $\eta$, that is, $t'_{acc} = R_L/\eta c$, where $R_L$ is the Larmor radius of electrons and $\eta < 1$. The total cooling timescale, $t'_{cool}$, included synchrotron \citep{Rybic86}, IC \citep{Aharo81}, and adiabatic cooling processes \citep{Zdzia14}.

The evolution in time of the electron distribution within the jet is described with the kinetic equation,
\begin{equation}
\frac{\partial N_e'(\gamma, t')}{\partial t'} = \frac{\partial }{\partial \gamma}\left ( \frac{\dd \gamma_{tot}}{\dd t'} N_e'(\gamma, t') \right ) + Q_{inj}(\gamma, t'),
\label{elec-eq}
\end{equation}
where $N_e'$ is the differential spectrum of electrons, and $\dd \gamma_{tot}/\dd t' = \gamma/t'_{cool}$ is the total cooling rate. The kinetic equation was solved numerically with the method presented in \citet{Chiab99}. Fresh electrons were injected into the jet up to $z_{end} = \Gamma_j \beta_j c t_{end}'$. However, the calculations were extended to $z_{max} = 10 \times z_{end}$ to ensure that the electrons were efficiently cooled within the jet.  

To derive the cooling rate, $\dd \gamma_{tot}/\dd t' = \gamma/t'_{cool}$, information is needed on the local radiation field at a specific location within the jet because we included the IC cooling process in this calculations. For this reason, we adopted the generation method to calculate the steady distribution of electrons
in a specific region of the jet \citep[see details in][]{Banas18}. In the first approximation, we assumed that only the synchrotron and the adiabatic cooling processes exist within the jet. In the next step, when the soft radiation field was known, we solved the kinetic equation, also including the IC process. We repeated this procedure a few times until a stable solution of the electron spectrum was obtained as a function of the distance from the base of the jet. Our calculation shows that when this procedure is repeated three times, we obtain a stable spectrum of electrons. Further repetitions of the procedure do not change the results significantly. 

To obtain the local density of the low-energy radiation, $n'_{ph}$, at a specific distance $z'=\beta_j c t'$, we need to take into account the radiation produced in the whole jet. For simplicity, we assumed that photons arriving at the jet layer at a specific distance, $z'$, have only two possible directions. In this way, we approximated the real angular distribution of photons produced in other regions of the jet by two monodirectional beams. This simplification of the angular distribution of the seed photons is justified by the very elongated structure of the jet, whose half-opening angle was assumed to be equal to 0.033 rad. This means that the jet is 30 times longer than it is wide. Thus, the direction of the seed photons that have reached the electrons is highly anisotropic.

The direction of photons produced in the bottom regions of the jet was assumed to be consistent with the jet axis (we call them backward photons and represent them by the subscript '$+$'). On the other hand, the direction of photons produced at greater distances from the base of the jet than the location of the electrons is opposite to the direction of the jet axis. They are called forward photons (represented by the subscript '$-$'). This simplification is justified by the elongated shape of the jet,  $(z_{end} - z_{0}) \gg R_j$. In this case, the local radiation in the specific location in the jet is approximated by photons from the forward or backward regions of the jet. We determined the local density of photons, $n_{\pm}(\epsilon', t')$, by solving the radiation transfer equation,

\begin{equation}
 n'_{-}(\epsilon'; t') = \frac{1}{c \epsilon'} \int_{z'}^{z'_{max}} j'(\epsilon', \Omega; z_{-}) \exp(-\tau_{SSA}) \Delta \Omega \dd z_{-},
 \label{n_ph_minus}
 \end{equation}
\begin{equation}
 n'_{+}(\epsilon'; t') = \frac{1}{c \epsilon'} \int_{z_0}^{z'} j'(\epsilon', \Omega; z_{+}) \exp(-\tau_{SSA}) \Delta \Omega \dd z_{+},
 \label{n_ph_plus}
\end{equation}
where $j(\epsilon', \Omega; z_\pm)$ is the synchrotron emissivity at the distance $z_\pm$ , and $\tau_{SSA}$ is the optical depth for the self-synchrotron absorption (SSA) process. $\Delta\Omega$ is the solid angle of incoming radiation, given by
\begin{equation}
  \Delta\Omega(z', z_{\pm}) = \frac{2 \pi r h}{r^2} = 2 \pi ( 1 - \frac{|(z_{\pm} - z')|}{r}),
\end{equation}
where $2 \pi r h$ is the area of the spherical layer, and $r = [(z' - z_{\pm})^2 + R_j(z')^2]^{1/2}$ and $h = r - (z' - z_{\pm})$.

The knowledge of the local density of soft radiation and the differential spectrum of electrons allows us to calculate the emissivities ($j$) and absorption coefficients ($\alpha$) for the synchrotron, IC, synchrotron self-absorption, and $\gamma-\gamma \rightarrow  \ e^{\pm}$ pair creation processes in the plasma rest frame at an arbitrary location within the jet. Then, we obtain the intensity of the emission from the jet by solving the equation of radiative transfer \citep{Rybic86},

\begin{equation}
 I_{jet} = \int_{z_0}^{z_{max}} j(z) \times \exp\left(-\int_z^{z_{max}} \alpha(z^*) \dd z^*\right) \dd z.
\end{equation}

\noindent This equation is true only when the observation angle is smaller than or equal to the half-opening angle of the jet. This equation can therefore only be applied to blazars for which the inclination angle of the jet is smaller than the jet half-opening angle. Knowledge of the intensity of the emission from specific regions of the jet allows us to calculate the total spectral energy distribution of radiation produced in the jet. 

\subsection{Intrinsic absorption of $\gamma$-rays within the jet}

In the previous section, we calculated the radiation structure of the jet. Here we consider the fate of $\gamma$-ray photons, which, if injected close to the base of the jet, have to propagate in this inhomogeneous radiation field.
The optical depth for the absorption of the $\gamma$-rays in collisions with the synchrotron radiation from the jet (i.e., $\gamma + \gamma \rightarrow  \ e^{\pm}$), $\tau_{\gamma\gamma}$, is proportional to the product of the photon density and the radius of the emission region, that is, $\tau_{\gamma\gamma} \propto n_{ph}r_{em}$. The exact mathematical description of this process may be found in \citet{Gould67}. In terms of the simple homogeneous SSC models, the geometry of the emission region is approximated as a sphere with radius $r_{em} \approx 0.5 c D t_{var}$ \citep[see, e.g.,][]{Bedna97c}. For these conditions, the escape probability of $\gamma$-rays is
\begin{equation}
p_{esc} = \frac{1-e^{-\tau_{\gamma\gamma}}}{\tau_{\gamma\gamma}}.
\end{equation}
\noindent
The situation is much more complicated when $\gamma$-rays propagate in the radiation field of an inhomogeneous jet. For such a jet, the optical depth depends on the direction and the emission place of the $\gamma$-rays. If the $\gamma$-ray photon is produced inside the jet and propagates along the jet axis, the optical depth, $\tau_{\gamma\gamma}$, can be calculated by integrating the local absorption coefficient along the jet axis,
\begin{equation}
\tau_{\gamma\gamma}(\epsilon; z_{em}) = \int_{z_{em}}^{z_{max}} \alpha_{\gamma\gamma}(\epsilon, z) \dd z,
\end{equation}
\noindent where $z_{max}$ is the end of the active part of the jet, and $\alpha_{\gamma\gamma}$ is the absorption coefficient. For small observation angles, as in the case of blazars, the path of the $\gamma$-rays in the radiation field of the jet is much longer than in typical SSC models. On the other hand, the target radiation is produced in the whole (very extended) volume of the parsec-scale jet, but not only within the compact blob. Because of this, the density of the low-energy photons is much lower.

\section{Application of the nonlocal inhomogeneous stationary jet model to Mrk 421}

We applied the model of the stationary jet described above to the broadband spectrum of the BL Lac object Mrk 421. This blazar was observed in 2013 in the low-activity state by \citet{Balok16} across the whole range of the electromagnetic spectrum. This persistent nonflaring state is naturally explained in terms of our stationary jet model.
In our modeling, we assumed that the mass of the SMBH is equal to $M_{BH} = 3 \times 10^8 M_\odot$ , where $M_\odot$ is the mass of the Sun. This value agrees with the mass estimated for the SMBH in Mrk 421 \citep[see][]{Wu02, Barth03}. The mass of the SMBH allowed us to fix the radius of the jet at the base assumed to be equal to 10 Schwarzschild radii,  $R_0 = 10 R_{Sch}$. The parameters that define the geometry of the jet, the half-opening angle of the jet, $\phi$, were chosen based on radio observations of the parsec-scale jet \citep{Claus13}, that is, it is equal to $\phi=0.2/\Gamma_j$ rad. 

\begin{figure}
        \resizebox{\hsize}{!}{\includegraphics{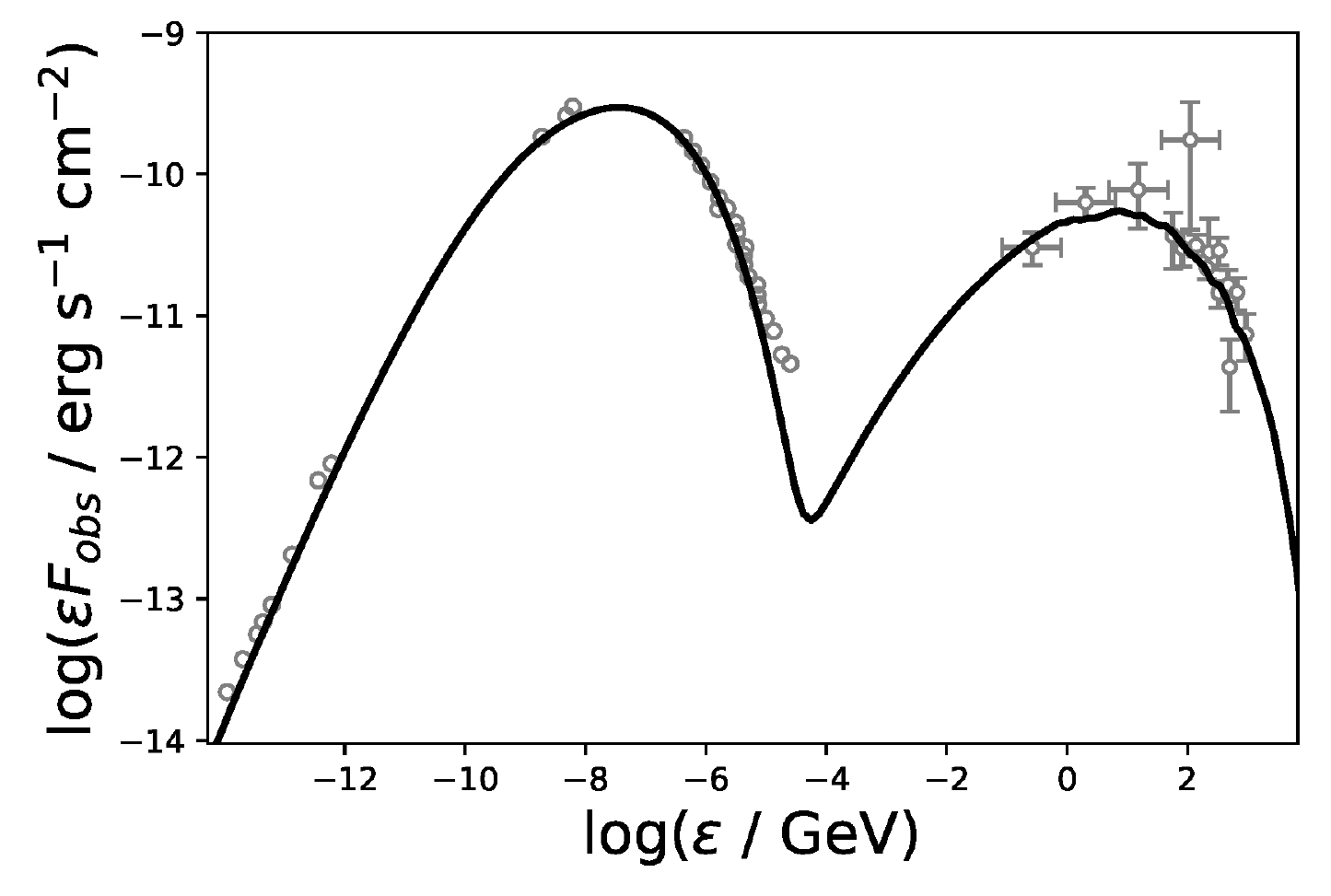}}
        \caption{Spectral energy distribution of Mrk 421 during the observations at 56302 MJD; data are from \cite{Balok16} (see circles). The solid black line is the fit to the differential spectrum obtained in terms of the nonlocal, inhomogeneous jet model. The parameters of the model are listed in Table \ref{stationary-parameters}.}
  \label{fig-sed}
\end{figure}

Based on the radio observations of parsec-scale jets \citep{Lico12}, we selected the observation angle $\theta_{obs}$ and the bulk Lorentz factor $\Gamma_j$. In this modeling, we assumed that the bulk Lorentz factor of the jet was relatively low, that is, $\Gamma_j = 6$. The jet is seen at an angle, $\theta_{obs} = 0.033$ rad ($\sim 2^\circ$). For these jet parameters, the observation angle is equal to the half-opening angle of the jet. This means that $\gamma$-rays propagate toward the observer within the radiation field of the jet up to the end of its active region, which is defined by 
$z_{max}$. 
The other parameters of our jet model were selected based on fitting the calculated spectral energy distribution to the observed broadband spectrum of Mrk 421. These parameters are reported in Tab.~\ref{stationary-parameters}. The calculated synchrotron and the IC spectra are compared with the observations in Fig.~\ref{fig-sed}. The calculated spectrum fits the observations in every range of the electromagnetic spectrum well, starting from the radio range up to VHE $\gamma$-ray energies.   

\begin{table}
\caption{Jet parameters for the stationary and inhomogeneous model of Mrk 421.}
\label{stationary-parameters}
\centering
\resizebox{\columnwidth}{!}{%
\begin{tabular}{lccc}
\hline \hline
Parameter & Symbol & Value &  Unit  \\
\hline
Mass of the black hole & $M_{BH}$ & $3 \times 10^8$ & $M_\odot$ \\
Bulk Lorentz factor of the jet & $\Gamma_j$ & $6$ & $-$ \\
Radius of the jet at the base & $R_0$ & $10$ & $R_{Sch}$ \\
Half opening angle of the jet & $\phi$ & $0.033$ & rad \\
Magnetic field at the jet base & $B_0$ & $3.5$ & G \\
Power injected in electrons & $L'_{inj}$ & $1.1 \times 10^{42}$ & erg/s \\
Index of injection profile & $q$ & $0.8$ & - \\
Spectral index of electrons & $s$ & $1.5$ & - \\
Acceleration coefficient & $\eta$ & $1.5 \times 10^{-7}$ & - \\
Minimum energy of electrons & $\gamma_{min}$ & $2 \times 10^3$ & $m_e c^2$ \\
End point of electron acceleration & $z_{end}$ & $10^6$ & $R_{Sch}$ \\
Observation angle & $\theta_{obs}$ & $0.033$ & rad \\
\hline
\end{tabular}
}
\end{table}

\begin{figure}
  \resizebox{\hsize}{!}{\includegraphics{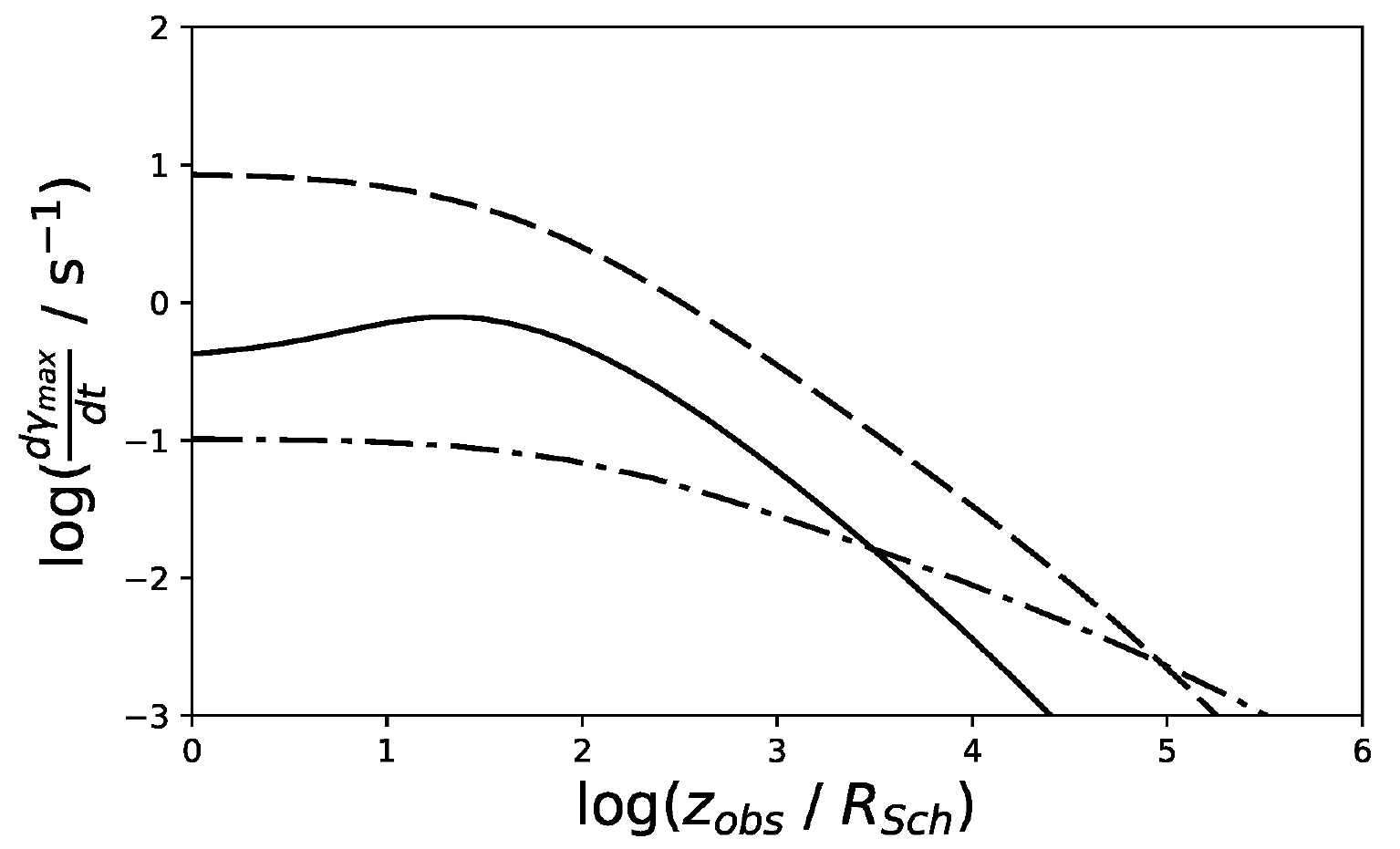}}
  \caption{Local loss rates in the plasma rest frame for three considered cooling processes:
synchrotron (dashed line), IC (solid), and adiabatic (dot-dashed) processes. All loss rates are calculated for electrons with the local maximum Lorentz factor, $\gamma_{max}$, obtained using a comparison between the acceleration timescale, $t'_{acc}$, and the total cooling timescale, $t'_{cool}$.}
  \label{fig-loss-rates}
\end{figure}
\noindent

\begin{figure}
  \resizebox{\hsize}{!}{\includegraphics{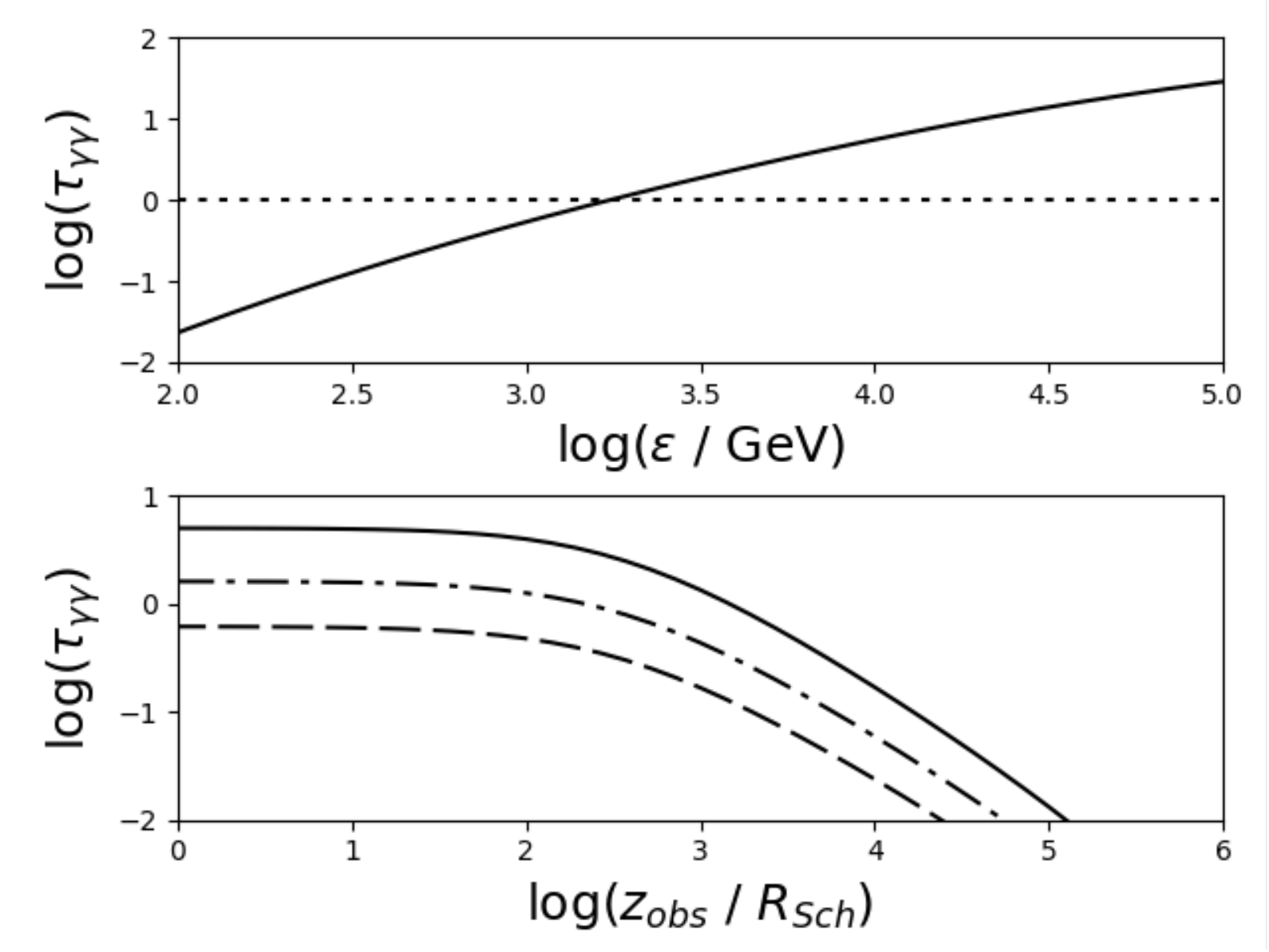}}
  \caption{Optical depth for the $\gamma + \gamma \rightarrow e^+ + e^-$ process, calculated in the nonlocal inhomogeneous jet model for the parameters reported in Tab.~\ref{stationary-parameters} (upper panel). Top: Optical depth as a function of the $\gamma$-ray photon energy. The opacity threshold 
($\tau_{\gamma-\gamma} = 1$) is shown with the dashed line. Bottom: Optical depth for the $\gamma + \gamma \rightarrow e^+ + e^-$ process as a function of distance from the jet base for a few selected energies of $\gamma$-ray photons: $\epsilon_\gamma = 1$ TeV (dashed line),  $\epsilon_\gamma = 3$ TeV (dot-dashed line), and $\epsilon_\gamma = 10$ TeV (solid line; bottom panel).}
  \label{fig-tau}
\end{figure}
\noindent

The presented spectrum shows the cumulated radiation from the whole volume of the jet. In the case of the inhomogeneous jet model, the conditions in the jet change along the jet axis, therefore the ratio of different cooling processes may change significantly. In Fig.~\ref{fig-loss-rates} we present the loss rates for three considered cooling processes: synchrotron, IC, and adiabatic processes. These loss rates were calculated for electrons with the highest local energy, which is a result of the balance between acceleration and cooling processes. We note that the relations between the different cooling processes change. At the beginning of the jet, most of the electron energy is emitted in the synchrotron process, which clearly dominates the other two cooling processes. With increasing distance from the jet base, the role of the IC cooling process increases. However, the cooling processes for the largest distances from the jet base are dominated by the adiabatic losses.

In this modeling, we also included the effect of absorption of $\gamma$-rays that are produced locally in the jet in the soft radiation field that is produced in other regions of the jet. We show that
the absorption process of VHE $\gamma$-rays is not negligible for the assumed parameters of the jet. The absorption becomes strong, that is, $\tau_{\gamma\gamma} = 1$, for VHE $\gamma$-rays with energies equal to $\epsilon_{tr} \simeq 2$ TeV. The optical depth for $\gamma$-rays that pass through the whole volume of the jet as a function of their energy is shown in Fig.~\ref{fig-tau}. The figure also shows the optical depth as a function of distance from the jet base. It becomes clear that the opacity of the jet decreases strongly for distances greater than $\sim 100 R_{Sch}$. $\gamma$-rays with energies of about 10 TeV can escape from the jet provided that they are produced at a distance larger than $\sim$1000 $R_{Sch}$. However, multi-TeV $\gamma$-rays, injected into the jet close to the jet base, are predicted to be absorbed efficiently.

\section{IC $e^\pm$ pair cascades initiated by primary gamma-rays within the jet}

A group of models for the $\gamma$-ray production in the vicinity of the SMBH is inspired by observations of exceptionally short flares from some AGNs. These events are characterized by a variability timescale close to or even shorter than the light-crossing time of the SMBH horizon. One of the most impressive examples was the $\gamma$-ray flare observed from IC 310 \citep{Aleks14}. The variability time was shorter than 5 minutes. This indicates that the stationary emission region should be a few times smaller than the Schwarzschild radius of the SMBH in IC 310 for the estimated values of the observation angles in the range between 0.17 and 0.35 rad \citep{Aleks14}. This suggests that the VHE $\gamma$-rays might be produced in the vicinity of SMBH. 

These observations of exceptionally short flares in the $\gamma$-ray range initiated research on the models in
which high-energy $\gamma$-rays are produced in the direct vicinity of the SMBH (see Introduction). In blazars, the line of sight of the jet axis can be smaller than the half-opening angle of the jet. In this case, the VHE $\gamma$-rays have to propagate from the vicinity of the SMBH within the soft radiation field of the jet before reaching the observer. In the case of opaque jets, $\gamma$-rays should be efficiently absorbed, initiating an IC $e^{\pm}$ pair cascade. In this section, we explore the details of the model in which the VHE $\gamma$-rays that are produced close to the SMBH  initiate these cascades inside the jet volume. 

In the case of blazars, the viewing angle of the jet is expected to be very small, in contrast to the radio galaxy IC 310. However, if similar phenomena occur in blazars as well, then the VHE $\gamma$-rays must cross the radiation field of the jet before reaching the observer. These  $\gamma$-rays can be absorbed in collisions with the nonthermal soft radiation from the jet itself. As a result of the absorption process, e$^{\pm}$ pairs are created. These e$^{\pm}$ pairs produce secondary photons in the IC process of low-energy radiation of the jet. They can also produce synchrotron radiation in the magnetic field of the jet. 
The energy of the created $e^{\pm}$ pairs depends on the energy of the primary VHE $\gamma$-rays, $\epsilon_0$. They can roughly be estimated as $\sim \epsilon_0/2$. The energies of secondary $e^\pm$ pairs do not depend directly on the local conditions within the jet that determine the energy losses of locally accelerated electrons. 
These secondary $e^\pm$ pairs are expected to have much higher energies that produce hard X-ray emission in the local magnetic field of the jet.

\begin{figure}
  \resizebox{\hsize}{!}{\includegraphics{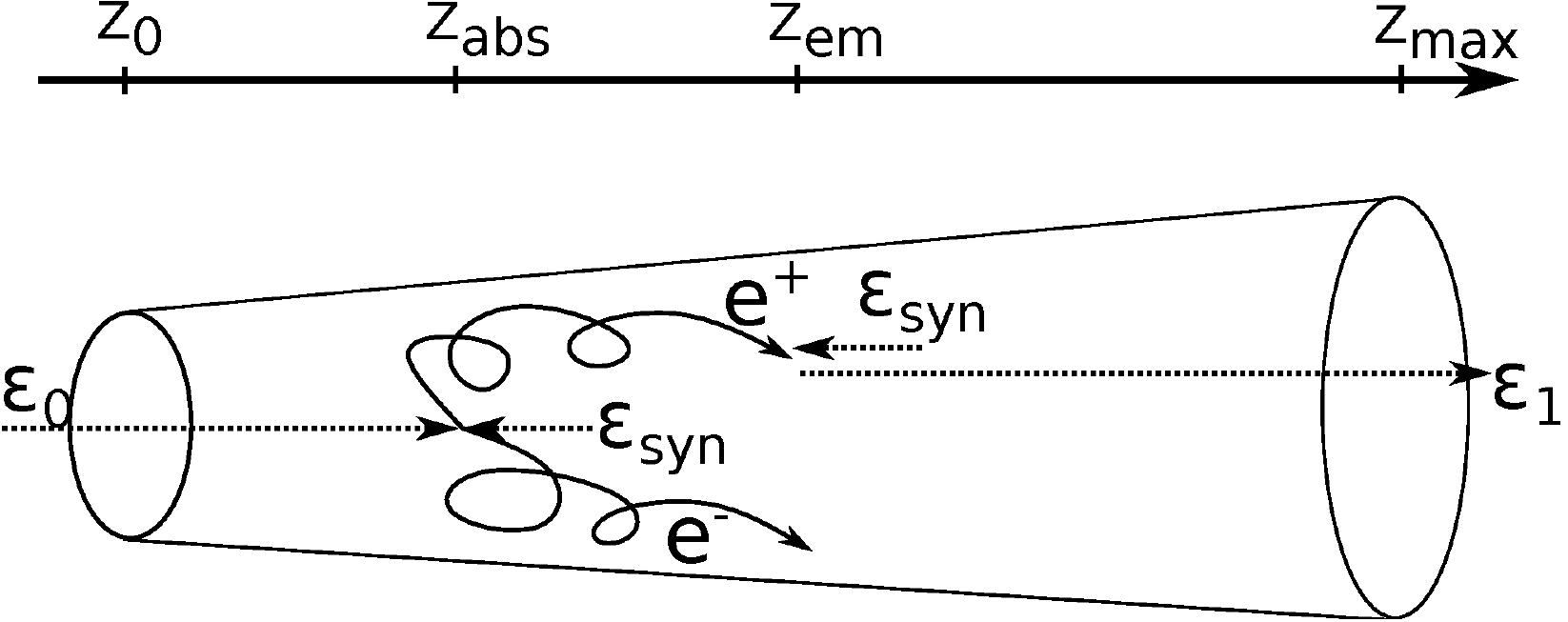}}
  \caption{Schematic picture of the IC $e^\pm$  cascade that develops within the jet volume (not to scale). The primary $\gamma$-ray photon, $\epsilon_0$, is injected at the jet base located at $z_0$. The photon is absorbed in the synchrotron radiation of the inhomogeneous jet. The secondary $e^\pm$ pair is created as a result of the absorption at a distance $z_{abs}$. This $e^\pm$ pair is isotropized in the random component of the magnetic field of the jet. It is rather advected with the jet plasma. The $e^\pm$ pair is cooled down, producing synchrotron radiation and secondary $\gamma$-ray photons in the IC process. These secondary photons, $\epsilon_1$, created at a distance $z_{em}$, can be also absorbed or escape from the jet.}
  \label{fig-cascade}
\end{figure}

In this section, we present the numerical model for the IC $e^\pm$ pair cascades initiated by VHE $\gamma$-rays in the radiation field of the stationary jet. We apply a scenario in which VHE $\gamma$-rays are produced in the vicinity of the SMBH. They propagate within the jet region. We apply the radiation field of the jet using the nonlocal inhomogeneous jet described in the previous section. In order to follow the IC $e^\pm$ pair cascades in the jet radiation, we use the dedicated Monte Carlo simulations discussed below. The general idea of our model is shown in Fig.~\ref{fig-cascade}.  

\noindent

\subsection{Basic assumptions and limitations of the numerical model}

The accurate calculation of the IC $e^\pm$ pair cascades that develop inside the jet volume is a very complex task. Because of this, we applied a number of assumptions mentioned below.

In reality, photons and $e^\pm$ in the electromagnetic cascades develop in three dimensions. In our model, we limit the problem to one dimension. We assumed that the primary VHE $\gamma$-rays move parallel to the jet axis. Similarly to the inhomogeneous model presented above, we assumed that the seed photons (in the comoving frame of the plasma) have only two possible directions: consistent with and opposite to the jet axis. This simplification is justified by the elongated shape of the jet, that is, the local radiation in such an extended jet is dominated by photons that are produced within the forward and backward regions of the jet.

We also assumed that the direction of the emitted secondary $\gamma$-ray photons is consistent with the direction of the jet axis. In reality, the secondary $\gamma$-rays are emitted into a cone with the opening angle equal to 1/$\Gamma_j$. Because of this, the real path of the secondary $\gamma$-rays in the radiation field of the jet may be slightly longer, but this should not affect the results noticeably.

The absorbed photons produce $e^{\pm}$ pairs inside the jet volume. We assumed that these $e^{\pm}$ pairs are immediately isotropized by the random component of the magnetic field (in the comoving frame of the plasma). The $e^{\pm}$ pairs are advected with the plasma along the jet axis. 
  
The synchrotron and IC radiation emitted by the $e^{\pm}$ pairs might have an effect on the local conditions in the jet. In particular, the changes in the radiation field modify the local absorption coefficient of the $\gamma$-ray absorption. However, we did not include this effect in our calculations. 

The last simplification is that the secondary synchrotron and IC radiation is uniformly collimated into a cone with an opening angle equal to 1/$\Gamma_j$.

\subsection{Numerical implementation of the model}

We assumed that $\gamma$-rays are injected at the base of the jet with a power-law spectrum described by a single spectral index.
We simulated the energies of the injected $\gamma$-ray photons from this distribution. The selected $\gamma$-ray propagates in the extended jet 
radiation field. The absorption probability of this primary photon is given by
\begin{equation}
p_\gamma(\epsilon_0) = 1 - \exp(-\tau_{\gamma\gamma}(\epsilon_0),
\end{equation} 
\noindent 
where $\tau_{\gamma\gamma}$ is the optical depth (see Sect.~2.2).
The distance, $z_{abs}$,  at which the $\gamma$-ray is absorbed is calculated by using the inverse function of the cumulative 
distribution,
\begin{equation}
z_{abs} = F^{-1}(u),
\end{equation}
\noindent 
where $u$ is a random number simulated from a uniform distribution over (0, 1), and $F$ is the cumulative distribution function 
given by
\begin{equation}
F(z_{abs}; \epsilon_0) = \frac{1 - \exp(-\int_{z_0}^{z_{abs}} \alpha_{\gamma\gamma}(z, \epsilon_0))\dd z}{1 - \exp(-\tau_{\gamma\gamma}(\epsilon_0))}.
\end{equation}
To determine the energies of the created $e^{\pm}$ pairs, we applied the procedure described in \citet{Boett97}. 
At first, the photon energies were transferred to the center-of-momentum frame. In this frame, the $e^{\pm}$ pair was created with 
equal energies and with the opposite momentum vector. Then, the pair energies were transformed into the reference frame of the jet. 
The created $e^{\pm}$ pairs were iteratively cooled down in the synchrotron and the IC processes. The energies of $e^\pm$ pairs were continuously reduced due to the synchrotron energy losses. The procedure for the IC process differs significantly because $e^\pm$
suffer discrete energy losses during a single interaction because the scattering occurs in the 
Klein-Nishina regime \citep{Blume70}. The secondary $\gamma$-rays that are produced can be farther absorbed in the radiation field of the extended 
jet. We followed the process of the cooling of the cascade $e^\pm$ pairs to the moment they left the volume of the jet or their energies decreased below 100 MeV.

\section{Application of the cascade inside the jet model to blazars}

We applied the cascade inside the jet model in order to explain the curious feature in the X-ray spectrum of blazar Mrk 421. 
\citet{Katao16} observed a concave shape of the X-ray spectrum during the low-flux state. 
The authors suggested that this excess is the low-energy tail of the IC bump. However, This explanation may be problematic for 
several reasons. (1) The estimated upper limit of the variability timescale is quite short in comparison to the radiative cooling 
timescale \citep{Katao16}. (2) The radio emission from low-energy electrons exceeds the observed radio flux even when the synchrotron self-absorption effect is taken into account \citep{Chen17}. (3) The other observations \citep{Accia21} also observed 
the hard X-ray excess, but it was clearly above the extrapolation of the average IC $\gamma$-ray spectrum.
Because of these arguments, another explanation of the hard x-ray emission from this source is required.

It is important to note that the observation of the hard X-ray excess in the case of hard-spectrum blazars (HBL blazars) 
has also been pointed out by
other authors. A similar concave shape in the X-ray range has been observed before by \citet{Madej16} in the case of PKS 2155-304. 
Moreover, other observations of Mrk 421, again in the low state, observed an excess in the hard X-ray energy range \citep{Accia21}. 
The existence of such a hard X-ray feature can be explained in terms of the multicomponent model, in which emission
with different properties is related to the presence of the spine-shear structure of the jet \citep{Chen17}. 

\begin{figure}
  \resizebox{\hsize}{!}{\includegraphics{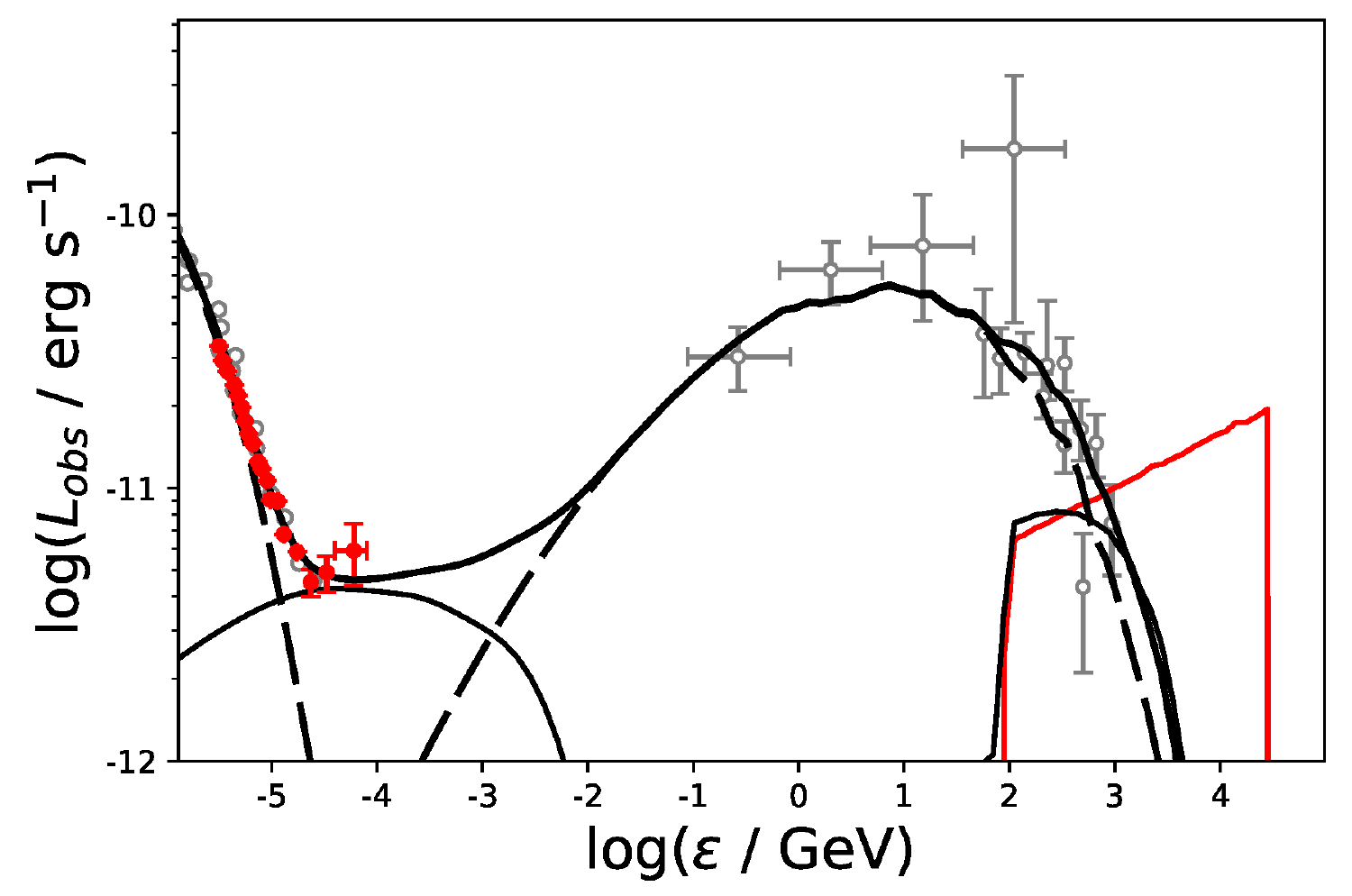}}
  \caption{X-ray and $\gamma$-ray spectrum of Mrk 421 with the visible excess at hard X-rays (red circles with error bars; data from \citet{Katao16}). The primary $\gamma$-rays are injected into the jet (solid red line) with the spectrum observed in the misaligned blazar IC 310 \citep{Aleks14}. The secondary emission from the cascades is shown with the thin black line (in the X-ray and HE $\gamma$-ray range). The emission from the stationary jet is presented with the thick dashed line. The total emission from the jet is represented by the thick solid line. The gray circles show the spectral energy distribution of Mrk 421 observed at 56302 MJD by \citet{Balok16}. The parameters of the extended jet model are reported in Tab.~1.}
  \label{fig-sed-casc}
\end{figure}

However, another possible explanation for this hard X-ray is provided by our cascade in the jet radiation model.
For the known conditions inside the jet (see Section 3), we applied our model of the inhomogeneous cascade in the jet radiation model. 
We assumed that the features of the primary VHE $\gamma$-ray emission from the vicinity of the SMBH agree well
with the features of the $\gamma$-ray spectrum observed from the short-scale flare detected from the radio galaxy 
IC 310 \citep{Aleks14}. Then, we applied the power-law spectrum for the primary $\gamma$-rays with a spectral index equal to 
$s = 1.8$. The lower and upper cutofsf of the spectrum were set to $\epsilon_{min} = 100$ GeV and $\epsilon_{max} = 30$ TeV, 
respectively. This primary spectrum is shown in Fig.~\ref{fig-sed-casc} with the solid red line. 

The VHE $\gamma$-rays from the vicinity of the SMBH have to propagate within the volume of the extended jet in the case of 
blazars such as Mrk 421. They can be absorbed in the radiation field of the jet and produce 
the IC $e^\pm$ pair cascades. Our simulations show that these primary $\gamma$-rays should be efficiently 
absorbed in the extended jet. However, the contribution of this cascade component is not significant for the $\gamma$-ray spectrum
that is produced in the extended jet. On the other hand, the new spectral component that formed in the cascade process of primary $\gamma$-rays 
in the vicinity of the SMBH appears in the hard X-ray energy range. This is the result of the synchrotron emission from the secondary 
$e^\pm$ pairs of the  
cascade. The variability time of this radiation mainly depends on the injection time of the VHE $\gamma$-rays that are produced 
in the vicinity of SMBH. This is the result of the small observation angle and very short cooling time of secondary VHE $e^\pm$ pairs in 
the volume of the extended jet.
The energy range and shape of this hard X-ray component are consistent with the observed X-ray excess in Mrk 421 
by \citet{Katao16}. In our opinion, 
it is a natural explanation of the concave shape of X-ray radiation observed not only in Mrk 421, but also in PKS 2155-304. 
The future observation of BL Lacs objects in hard X-rays should verify the explanation we discussed here for this curious spectral feature.

\noindent

\section{Conclusions}

The main result of our work is the explanation of the excess of the hard X-ray emission observed in Mrk 421. To do this, we developed the nonlocal, inhomogeneous model for the stationary jet in active galaxies as a first step. 
The model is characterized by a relatively small number of free parameters. It was successfully applied to explain the multiwavelength spectrum of Mrk 421 in the nonflaring state. Next, we showed that the extended jet in blazars 
can be opaque for $\gamma$-rays with the highest energies. 
This happens in the case of the Mrk 421 if a relatively low bulk Lorentz factor of the jet is considered. 
This low value corresponds 
much better to the value estimated based on the radio observation of jets in radio-loud AGNs, which  
indicates that most of these types of objects are characterized by the Lorentz factors in the range from 5 to 15 \citep{Jorst17, Liste19}. 

In the second step, we considered the fate of VHE $\gamma$-rays, which are produced close to the vicinity of the 
SMBH and propagate through the extended jet radiation toward the observer. In the case of Mrk 421, we obtained that
the optical depth for these $\gamma$-rays is higher than one for $\gamma$-rays with energies higher than 2 TeV.
Nevertheless, the jet becomes transparent even for multi-TeV $\gamma$-rays at distances exceeding $100 R_{Sch}$

The radiation field produced in terms of the inhomogeneous jet model was used to calculate the radiation from 
the IC $e^\pm$ pair cascade that developed by $\gamma$-rays inside the extended jet. We investigated the consequences of the hypothesis in which the VHE $\gamma$-ray flare in the vicinity of the SMBH, similar to that from IC 310, might also appear in blazars in which jets are viewed at a very small angle. We showed that the consequence of this assumption is the appearance of the radiation feature of the secondary synchrotron emission in the hard X-ray part of the spectrum. The appearance of this emission is a natural explanation of the peculiar X-ray excess that is observed in the spectrum of Mrk 421 during the low state \citep{Katao16}. A similar feature has also been observed in another well-known blazar, PKS 2155-304 \citep{Madej16}. In contrast to the explanation proposing that this feature is due to an extension of the IC emission to low energies, our solution indicates that this X-ray feature could be highly variable. Its variability timescale could be marginally longer than the injection timescale of the primary VHE $\gamma$-rays due to the very short cooling time of secondary high-energy electrons and the very small observation angle of the jet.

\begin{acknowledgements}
This research is supported by the grants from the Polish National Research Centre No. 2015/19/N/ST9/01727 (PB+WB) and partially No. 2019/33/B/ST9/01904 (WB).
\end{acknowledgements}

\end{document}